\documentclass[
  twocolumn,
  showpacs,
  preprintnumbers,
  amsmath,
  amssymb,
  aps,
  prl,
  floatfix,
  letterpaper,
  superscriptaddress,
  nofootinbib
]{revtex4-2}
\usepackage{makecell}
\usepackage{feynmp-auto}
\usepackage[utf8]{inputenc}
\usepackage{xcolor}
\makeatletter
\def\pdfstartlink@attr{}
\makeatother
\usepackage{tabularx}
\usepackage{graphicx}
\usepackage{dcolumn}     
\usepackage{multirow}
\usepackage{bm}          
\usepackage{hyperref}

\usepackage{parskip}
\def\fmfovalblob#1#2#3{\fmfcmd{input vovalblob; vovalblob ((#1), (#2), \fmfpfx{#3});}}
\def\fmfovalblobz#1#2#3{\fmfcmd{input vovalblob; vovalblobz ((#1), (#2), \fmfpfx{#3});}}

\definecolor{green}{rgb}{0,.6,0}

\def\today{\ifcase\month\or
  January\or February\or March\or April\or May\or June\or
  July\or August\or September\or October\or November\or December\fi
  \space\number\day, \number\year}

\newcommand\UNH{University of New Hampshire, Durham, New Hampshire 03824, USA}
\newcommand\UVA{University of Virginia, Charlottesville, Virginia 22903, USA}
\newcommand\WM{William \& Mary, Williamsburg, Virginia 23187, USA}

\newcommand\Jlab{Thomas Jefferson National Accelerator Facility,
  Newport News, Virginia 23606, USA }

\newcommand\INFNGE{Istituto Nazionale di Fisica Nucleare, Sezione di Genova, I-16146 Genova, Italy}

\newcommand\ODU{Old Dominion University, Norfolk, Virginia 23529, USA}

\newcommand\Duke{Duke University, Durham, NC 27708, USA}

\newcommand\Mainz{Institute of Nuclear Physics, Johannes Gutenberg Universit$\ddot{a}$t Mainz, 55099 Mainz, Germany}
\newcommand\MainzPrisma{PRISMA$^+$\,Cluster of Excellence, Johannes Gutenberg Universit$\ddot{a}$t Mainz, 55099 Mainz, Germany}
\newcommand\Scherrer{Laboratory for Particle Physics, Paul Scherrer Institute, 5232 Villigen PSI, Switzerland}

\include{review_tool.tex}

\begin{document}

\title{New Spin Structure Constraints on Hyperfine Splitting and Proton Size}
\author{D. Ruth} \thanks{Corresponding Author, Email: david.ruth@unh.edu}\affiliation{\UNH}
\author{K.~Slifer} \affiliation{\UNH}
\author{J.-P.~Chen} \affiliation{\Jlab}
\author{C.~E.~Carlson} \affiliation{\WM}
\author{F. Hagelstein} \affiliation{\Mainz}\affiliation{\MainzPrisma}\affiliation{\Scherrer}
\author{V. Pascalutsa} \affiliation{\Mainz}
\author{A. Deur} \affiliation{\Jlab}
\author{S. Kuhn} \affiliation{\ODU}
\author{M. Ripani} \affiliation{\INFNGE}
\author{X. Zheng} \affiliation{\UVA}
\author{R. Zielinski} \affiliation{\UNH}
\author{C. Gu} \affiliation{\Duke}

\noaffiliation

\date{\today}

\begin{abstract}

The 1S hyperfine splitting in hydrogen is measured to an impressive ppt precision and will soon be measured to ppm precision in muonic hydrogen. The latter measurement will rely on theoretical predictions, which are limited by knowledge of the proton polarizability effect $\Delta_\text{pol}$. Data-driven evaluations of $\Delta_\text{pol}$ have long been in significant tension with baryon chiral perturbation theory. Here we present improved results for $\Delta_\text{pol}$ driven by new spin structure data, reducing the long-standing tension between theory and experiment and halving the dominating uncertainty in hyperfine splitting calculations.

\end{abstract}


\maketitle


The hyperfine splitting in hydrogen, the renowned 21 cm line arising from the magnetic dipole interaction of electron and proton, stands out as one of the best-measured quantities in physics, currently known to an impressive 12 digits.
This level of precision is challenging for theory to match, particularly in accounting for the effects of proton structure~\cite{Hyperfine}. These structure effects are amplified in muonic hydrogen, where the hydrogen's electron is replaced by a muon. 
Presently, several high-profile experiments~\cite{Amaro:2021goz,Pizzolotto:2020fue} are aiming at a first-ever measurement of the ground-state hyperfine splitting in muonic hydrogen. Their success in finding this forbiddingly narrow transition crucially depends on an accurate assessment of proton structure effects. However, previous determinations~\cite{Hyperfine2,Tomalak,Faustov} of the leading uncertainty among these proton structure effects, the proton polarizability effect $\Delta_\text{pol}$, have large error bars and are in significant tension with corresponding theoretical calculations~\cite{Hyperfine2,hyperfine_22,Hagelstein2023}. In this letter, we improve upon the evaluation of the proton polarizability contribution with new experimental proton spin structure data from Thomas Jefferson National Laboratory, significantly reducing this long-standing tension and halving this quantity's uncertainty.

The classical picture of the hydrogen atom is fairly simple: a pointlike spinless electron bound by a pointlike spinless proton via the Coulomb force. The Schr\"odinger equation gives the energy spectrum in natural units as $E_n=-\alpha^2 m_r/(2 n^2)$, with $n$ the principal quantum number and $m_r$ the reduced mass. The more sophisticated picture of the atom is modeled as a correction to this simple picture, including the effects due to spin and the structure of the proton, to be discussed here. The proton structure effects are small, but clearly seen
in the hydrogen spectrum at the current level of precision. They are more prominent in muonic hydrogen, because of a much smaller Bohr radius, $a_B = 1/(\alpha m_r)$, given that $m_r$ goes roughly as the lepton mass, which is 200 times heavier for the muon. 
The muon has thus 200$^3$ greater probability [as given by the  wave-function squared at the origin, $\left|\Psi_{n} (0)\right|^2 =1/ (\pi a_B^3 n^3)$] to be probing the proton substructure.


Because of this heightened sensitivity to nuclear structure details in muonic atoms, the recent breakthrough in the laser spectroscopy of 
muonic hydrogen ($\mu$H) by the CREMA Collaboration led to an order-of-magnitude
improvement in the measurement of the proton charge radius \cite{Pohl:2010zza}. Surprisingly
to many, it appeared 
to be $7\,\sigma$
smaller than the most-recent CODATA recommended value of the charge radius at the time \cite{Mohr:2012aa}.
This spectacular discrepancy, dubbed as the ``proton radius puzzle" (see \cite{Carlson:2015jba,Pohl:2013yb} for an early review) has since been
largely resolved and, from 2018 onwards, CODATA recommends the 
smaller, and more precise, $\mu$H value.
This chapter is not yet closed, with many new measurements of the proton charge radius 
underway using the conventional
methods of normal hydrogen (H) spectroscopy, elastic electron-proton ($ep$) 
scattering and even muon-proton ($\mu p$) scattering, see \cite{hyperfine_22,Karr:2020wgh,Gao:2021sml} for recent reviews. 
Here we concern ourselves with the calculation of the $\mu$H hyperfine splitting (HFS), and its implications for the next milestone of $\mu$H spectroscopy: the upcoming experimental measurement of the $\mu$H ground-state HFS.

Two different collaborations are competing to provide
this first-ever HFS measurement: CREMA~\cite{Amaro:2021goz} at the Paul Scherrer Institut (PSI) and FAMU~\cite{Pizzolotto:2020fue} at the RIKEN-RAL Muon Facility. Given the extreme narrowness of this transition, their success depends in part on how well the proton-structure corrections are understood, 
since the searches can only be done over a very limited range of frequencies.

To leading order, $\mathcal{O}(\alpha^4)$, the HFS is given by the Fermi energy
\begin{equation}
E_\mathrm{F}=\frac{8\alpha}{3a_B^3}\frac{1+\kappa_p}{m_l M_p},\label{FermiE}
\end{equation}
where $m_l$ is the lepton mass (either $m_e$ in H or $m_\mu$ in $\mu$H), and $M_p$ is the proton mass.  
Here the proton structure is only represented through the anomalous magnetic moment.

\begin{figure}[t]
\centering
\begin{fmffile}{tpe}
  \begin{fmfgraph*}(150,60)
    \fmfstraight
    \fmfleft{i2,K,i1}
    \fmfright{o2,Kbar,o1}
    \fmf{fermion,label.side=left,label=$e^-$/$\mu^-$}{i1,t1}
    \fmf{fermion,label.side=left,label=p}{i2,t4}
    \fmfshift{10 left}{K}
    \fmfovalblobz{.5w}{.4}{K}
    \fmfv{l.d=0,l.a=180,l=H/$\mu$H}{K}
    \fmf{phantom,tension=1}{t1,t2}
    \fmf{phantom,tension=0.5}{t4,t3}
    \fmf{phantom,tension=1}{t4,ti,t3}
    \fmf{fermion,tension=1,label.side=left,label=$e^-$/$\mu^-$}{t2,o1}
    \fmf{fermion,tension=1,label.side=left,label=p}{t3,o2}
    \fmfshift{10 right}{Kbar}
    \fmfovalblobz{.5w}{.4}{Kbar}
    \fmfv{l.d=0,l.a=3,l=H/$\mu$H}{Kbar}
    \fmf{fermion,tension=0,label.side=left}{t1,t2}
    \fmf{fermion,tension=0,label.side=left}{t4,t3}
    \fmfovalblob{.2w}{1.8}{ti}
    \fmf{boson,tension=0,label=$\gamma$,label.side=right}{t1,t4}
    \fmf{boson,tension=0,label=$\gamma$}{t3,t2}
  \end{fmfgraph*}
\end{fmffile}
\caption{Two-photon exchange diagram showing the interaction between electron ($e^-$) and proton (p) in hydrogen, or muon ($\mu^-$) and proton in muonic hydrogen.
}
\label{fig:epscattering}
\end{figure}
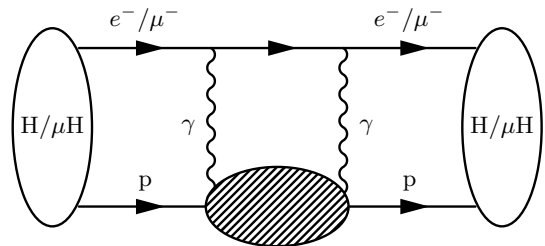

At the next order, $\mathcal{O}(\alpha^5)$, the proton structure effects can all be computed via the 
two-photon (2$\gamma$) exchange diagram of Fig.~\ref{fig:epscattering}, which usually is split into three
contributions:
\begin{align}
\label{hyperfine_2gamma}
E_{nS\text{-HFS}}^{2\gamma}= \frac{E_\mathrm{F}}{n^3}\big(\Delta_\mathrm{Z} + \Delta_\mathrm{recoil} + \Delta_\mathrm{pol}\big)
\end{align}

The largest, $\Delta_Z$, comes from the proton Zemach radius $R_Z$, a measure of how far the electric and magnetic distributions of the proton are correlated with each other. $R_Z$ is expressed in  terms of the elastic electric and magnetic form factors $G_E(Q^2)$ and $G_M(Q^2)$:
\begin{equation}
\label{RZdef}
R_{\mathrm{Z}}=-\frac{4}{\pi}\int_0^\infty \frac{d Q}{Q^2}\left[\frac{G_E(Q^2)G_M(Q^2)}{g_p}-1\right],
\end{equation}
with $g_p$ as the proton gyromagnetic g-factor. The recoil contribution $\Delta_\mathrm{recoil}$ can likewise be expressed in terms of the form factors, but the final contribution from the polarizability, $\Delta_\mathrm{pol}$, is more complicated. 

The polarizability effect is caused by the proton's moments induced
by the electromagnetic fields of the bound leptons. 
Unlike the radius, the polarizability contribution is not given by the form factors, but rather
by the inelastic structure functions $g_{1,2}(x,Q^2)$, which are functions of the Bjorken $x$, a variable which tracks the fraction of the interaction's momentum carried by one of the proton's quark constituents.  This contribution is more difficult
to obtain, due to the necessity to cover a 2-dimensional phase space, while the required spin structure function data are relatively sparse, especially at low $Q^2$ which dominates the determination. Previously, there had been only limited $g_1$ data and a complete lack of $g_2$ data in the kinematic region most relevant to the HFS. Nonetheless, a data-driven evaluation of this contribution has been attempted in the past \cite{Hyperfine2,Tomalak,Faustov}.

The present status is that the existing data-driven evaluations, while consistent with each other, are in disagreement with chiral perturbation theory ($\chi$PT), which predicts
a significantly smaller contribution of this effect~\cite{Hagelstein:2016yU,Hagelstein2023}.

Under
the general assumptions of unitarity (optical theorem) and analyticity (dispersion relations) 
of the forward Compton scattering, the contribution of the spin structure functions has the following 
form~\cite{Hyperfine2,hyperfine_22}:
\begin{eqnarray}
\label{deltapol}
\Delta_\mathrm{pol} &=& \frac{\alpha m_l}{2\pi (1+\kappa_p) M_p} \big(\Delta_1 + \Delta_2\big),\\
\label{eqn:hyperfine_delta1}
\Delta_1 &=&  \int\limits_0^\infty \frac{dQ^2}{Q^2}\bigg[\beta_1(\tau_l)  F_2^2(Q^2)
   \nonumber\\
&+&\frac{8 M_p^2}{Q^2}\int\limits_0^{x_\mathrm{th}}dx\,
\tilde{\beta}_1(\tau,\tau_l)\,g_1(x,Q^2)\bigg],\\
\label{eqn:hyperfine_delta2}
\Delta_2 &=& -24 M_p^2 \int\limits_0^\infty \frac{dQ^2}{Q^4} \int\limits_0^{x_\mathrm{th}}dx\,\tilde{\beta}_2(\tau,\tau_l)\,g_2(x,Q^2)
\end{eqnarray}
where $\beta$'s are elementary kinematic functions, $\tau$'s are kinematic variables, and $F_2$ is the Pauli form factor, the explicit definitions of which are in the Supplemental Materials.
$x_\mathrm{th}$ corresponds to the minimum energy necessary to generate a pion, at an invariant mass $W$ of 1073.2 MeV. 


\begin{figure}[t!]
\includegraphics[trim={0cm 0 1cm 1cm},clip,angle=0,width=0.9\columnwidth]{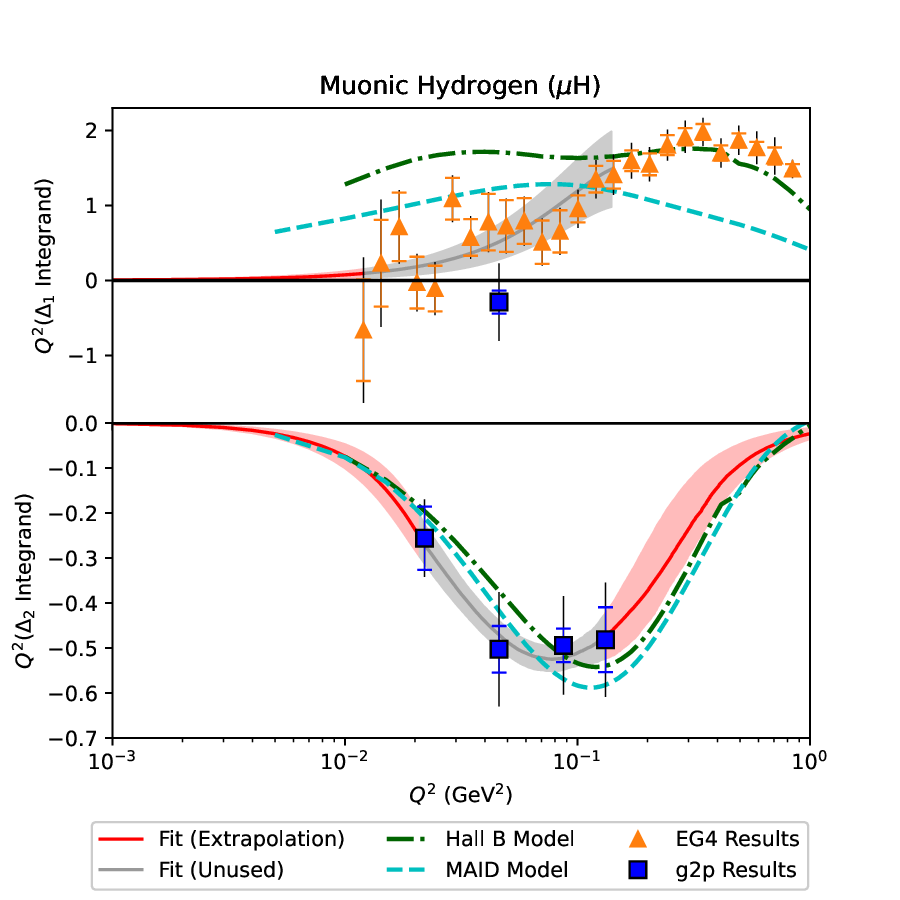}

\caption{The hyperfine contribution integrands for $\Delta_1$ and $\Delta_2$ in Eqs.~(\ref{eqn:hyperfine_delta1}) and (\ref{eqn:hyperfine_delta2}), weighted by $Q^2$, for muonic hydrogen. Results from the g2p experiment~\cite{g2p_nature} are shown in blue squares. The results of the EG4 experiment ~\cite{EG4_final} are shown in orange triangles. The inner error bars represent the statistical uncertainty, while the outer error bars represent the total uncertainty including systematic error. The green dash-dot and cyan dashed lines represent the phenomenological Hall B and MAID models~\cite{EG1b2,MAID2007} respectively. 
The form factor term of the integrand for $\Delta_1$ is constructed using the Arrington form factor fit~\cite{JAFit}. The red line indicates a new phenomenological fit to the data and extrapolation to low $Q^2$=0 and high $Q^2$, with the red band representing the uncertainty of the calculation. The results are similar but have different mass scaling in electronic hydrogen.
} 
\label{fig:hyperfine}
\end{figure}

$\Delta_\mathrm{pol}$ currently dominates the theoretical uncertainty of HFS calculations~\cite{hyperfine_22,Hyperfine} and it is evident that to calculate this contribution accurately, we must examine experimental measurements of the spin structure functions $g_1$ and $g_2$. Notably, the low-$Q^2$ regime dominates the integrals of  Eqs.~(\ref{eqn:hyperfine_delta1}) and (\ref{eqn:hyperfine_delta2}) due to the $\frac{1}{Q^4}$ factors, so it is especially vital to determine the spin structure functions at low $Q^2$ if we wish to fully understand the hydrogen atom, and by extension the HFS effect in general. 


\begin{figure}[t!]
\includegraphics[trim=0.1cm 0cm 0.1cm 0cm,clip=true,angle=0,width=0.9\columnwidth]{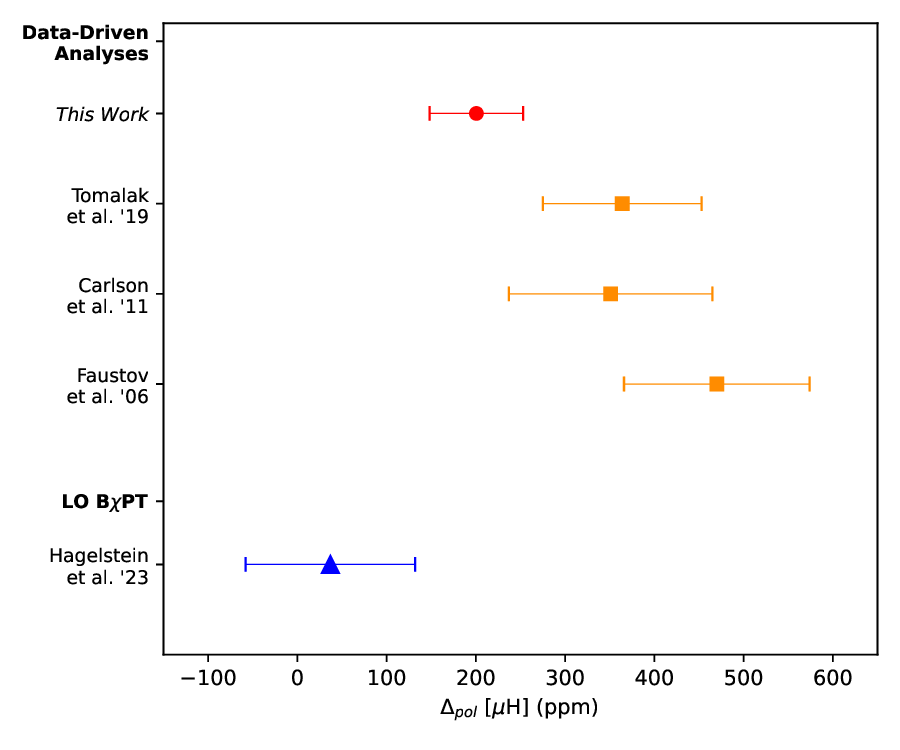}

\caption{
The polarizability contribution to the hyperfine splitting for muonic hydrogen. The analysis of this work is shown in a red circle, and is compared to previous data-driven dispersion relation calculations~\cite{Hyperfine2,Tomalak,Faustov} shown in orange squares, and the baryon chiral perturbation theory calculation~\cite{Hagelstein:2023owe} shown in blue triangles. 
} 
\label{fig:hcompare}
\end{figure}

In the following, we focus on a new empirical input for the proton spin structure functions. Our evaluation details new results from Jefferson Lab Experiments  E03-006 (EG4) and E08-027 (g2p), two complementary experiments both aimed at collecting low $Q^2$ data with longitudinally and transversely polarized proton (NH$_3$) targets, respectively~\cite{g2p_nature,EG4_final}. Here, we present these experiments' contribution to the hyperfine integrals above, with the g2p data providing the first data-driven extraction of $\Delta_2$, and the EG4 data providing new $\Delta_1$ data with unprecedented coverage in the low-$Q^2$ region.

In the EG4 experiment, 1-3.5 nA of longitudinally polarized electron beam was incident on a longitudinally polarized NH$_3$ target. The scattered electrons were detected using the CEBAF Large Acceptance Spectrometer (CLAS). The longitudinal polarized cross section difference $\Delta\sigma_\parallel$ was directly extracted from the yield difference between left- and right-handed beam electrons, such that contributions from the unpolarized material cancel. Combined with an estimation of the (small) transverse contribution based on a parameterization of world data, the proton structure function $g_1$ was extracted. Four different beam energies, along with the very small scattering angle down to 6$^\circ$, allowed the minimum $Q^2$ to reach a very low 0.012 GeV$^2$. These $g_1$ results were used to form the bulk of the low $Q^2$ $\Delta_1$ data presented in this letter, see Fig.~\ref{fig:hyperfine} (top panel).


In the g2p experiment, the parallel and perpendicular double spin asymmetries $A_\parallel$ and $A_\perp$ were measured for the scattering of 50 nA polarized electrons on longitudinally and transversely polarized NH${}_3$ targets, respectively. Scattered electrons were detected at an angle of $\approx$6.5$^\circ$ using the Hall A High Resolution Spectrometers and a Septa Magnet. Measured asymmetries were combined with unpolarized cross section models from the Bosted-Christy phenomenological fit~\cite{Bosted3} to form polarized cross section differences, which were used to extract the spin structure functions. By varying the polarized target magnetic field and electron beam energy, five different kinematic settings were measured ranging from $Q^2$ of 0.02 GeV$^2$ to 0.12 GeV$^2$. 
Four of these settings were measured with a transverse polarized target field, giving rise to a perpendicular polarized cross section difference and a $g_2$ result, and one setting with a longitudinally polarized target field, which provides a parallel polarized cross section difference and a $g_1$ result. The results from the g2p experiment are the first data in a range relevant to the HFS, and so are used to form the $\Delta_2$ results in this letter. 

Results for the $\Delta_1$ integrand are shown in the top of Fig.~\ref{fig:hyperfine}.
The unmeasured part of the integral, largely at low Bjorken-$x$, is estimated using the CLAS Hall B model~\cite{EG1b2}. This is the best available model, containing significantly more modern $g_1$ data than the Simula parametrization~\cite{Simula:2001iy} used in previous analyses~\cite{Hyperfine2}.
  A new phenomenological fit, shown in red, is generated to extrapolate to the low $Q^2$ region. 
  Details on the fitting procedure can be found in the Supplemental Materials. 
Numerical results for these contributions are obtained by integrating over the data where they exist, primarily the EG4 data shown~\cite{EG4_final}, as well as data from the EG1b experiment in the $Q^2$ = 1.0-5.0 GeV$^2$ region~\cite{EG1b2}. The contribution from the low-$Q^2$ regime is calculated by integrating the displayed extrapolation fit, while the high-$Q^2$ contribution above $Q^2$ = 5.0 GeV$^2$ is calculated using the Hall B Model~\cite{EG1b2}. 



Results for the  $\Delta_2$ integrand in Eq.~(\ref{eqn:hyperfine_delta2}) are shown in the bottom of Fig.~\ref{fig:hyperfine}. The unmeasured part of the $dx$ integral is again estimated using the Hall B model~\cite{EG1b2}. The results of g2p shown are the first ever direct experimental extractions of this quantity. The low and high $Q^2$ regions are calculated using the displayed fit, which is described in detail in the Supplemental Materials. 
Due to the comparative lack of $g_2$ data, the extrapolation has a somewhat larger error than for the $\Delta_1$ results.

This historical lack of $g_2$ data makes it difficult to conclude if the Hall B model~\cite{EG1b2} is a good estimation of the low-$x$ region or not. To account for this, we compare the result using the older Simula parametrization~\cite{Simula:2001iy}, which contains a significantly different prediction for the low-$x$ behaviour of $g_2$, and include the difference in our extrapolation error by comparing the upper and lower error bands of our extrapolating fit to the data in each case. Despite the very different models, this error contribution is relatively small, because the low-$x$ region is suppressed for $\Delta_2$. 

The integrated results for $\Delta_1$, $\Delta_2$, and $\Delta_\mathrm{pol}$ are as follows:

\begin{eqnarray}
\label{results_start}
\Delta_1^e &=& 6.78 \pm 1.02 (\text{data}) \pm 0.24(\text{extrapolation})\\
\Delta_1^\mu &=& 5.69 \pm 0.84 (\text{data}) \pm 0.20(\text{extrapolation})\\
\Delta_2^e &=& -1.98 \pm 0.16 (\text{data}) \pm 0.38(\text{extrapolation})\\
\label{results_end}
\Delta_2^\mu &=& -1.40 \pm 0.11 (\text{data}) \pm 0.31(\text{extrapolation})
\qquad
\end{eqnarray}
\begin{eqnarray}
\Delta_\text{pol}^e &=& 1.09 \text{ ppm}\pm 0.31\text{ ppm}\\
\Delta_\text{pol}^\mu &=& 200.6 \text{ ppm}\pm 52.4\text{ ppm}
\qquad
\end{eqnarray}

The total polarizability contribution to the hyperfine splitting is provided in parts per million (ppm) of the Fermi energy $E_F$. The uncertainties for $\Delta_1$ and $\Delta_2$ are divided into uncertainty coming directly from the data, and a combined systematic uncertainty coming from the extrapolations into high and low $Q^2$ regions and into the low-$x$ regime. The extrapolation error is calculated by generating pseudo-data within the data's error bars, and calculating a new fit to this pseudo-data. This procedure is repeated 1000 times, and the standard deviation in the resulting fits is taken as the extrapolation error band. The total extrapolation error in the table also includes a contribution from the choice of low-$x$ fill-in model, where the highest upper band and lowest lower band achievable with different choices of fill-in model are taken as the absolute limits of the error band. The data error is a combination of statistical and systematic error from the respective experiments contributing~\cite{g2p_nature,EG4_final,EG1b2}.


As can be seen in Fig.~\ref{fig:hcompare}, the new data from EG4 and g2p dramatically reduce the long
standing discrepancy between the leading-order (LO) $\chi$PT prediction~\cite{Hagelstein:2016yU,Hagelstein2023} and earlier data-driven dispersive evaluations~\cite{Hyperfine2,Tomalak,Faustov} of the polarizability contribution.
 The large difference from the earlier dispersive results 
 is illustrative of the importance of low-$Q^2$ data for $\Delta_\mathrm{pol}$, and the improvement in the available phenomenological models~\cite{EG1b2}, which are constrained by a larger amount of data compared to earlier parametrizations used in previous analyses~\cite{Simula:2001iy,Simula:2002tv}. 

\begin{figure}[t!]
\includegraphics[width=0.9\columnwidth]{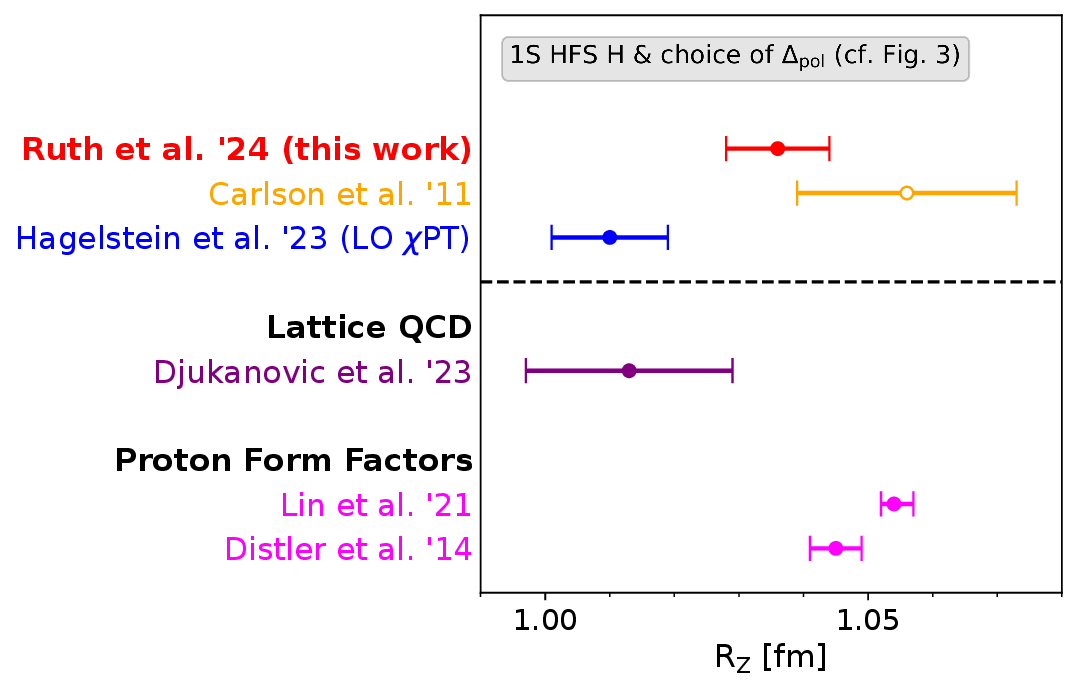}
\caption{Comparison of various extractions of the proton Zemach radius $R_\mathrm{Z}$ \cite{Hyperfine2,Hagelstein2023,Djukanovic:2023cqe,Lin:2021xrc,Distler:2010zq}.
}
\label{fig:RZ}
\end{figure}

Our new data-driven evaluations of $\Delta_\mathrm{pol}$ put us in the unique position to update the theoretical predictions of the HFS in (muonic) hydrogen, as well as the extractions of the proton Zemach radius from measurements of the HFS. The $1S$ HFS in H is extraordinarily well-measured 
\cite{Hellwig1970,Karshenboim:2000rg}:
\begin{equation}
E^{\,\text{exp.}}_{1S\text{-HFS}}(\mathrm{H})=1\,420.405\,751\,768(2)\,\text{MHz}.\label{expHHFS}
\end{equation}
Therefore, the presently most precise extraction of $R_\mathrm{Z}$ from spectroscopy is achieved when comparing the measured $1S$ HFS in H to the full theory prediction
including QED, electroweak and strong interaction effects:
\begin{equation}
E_{nS\text{-HFS}}=\frac{E_\mathrm{F}}{n^3}\left(1+\Delta_\mathrm{QED}+\Delta_\mathrm{weak}+\Delta_\mathrm{strong}\right).\label{eq:hfswoEF}
\end{equation}
For details on the numerical factors entering Eq.~(\ref{eq:hfswoEF}), we refer to the compilations in \cite[Eq.~(40) and (42)]{hyperfine_22}.
Here, $\Delta_\mathrm{strong}$ contains the 2$\gamma$-exchange contributions introduced in Eq.~(\ref{hyperfine_2gamma}), as well as other hadronic corrections such as hadronic vacuum polarization.
 The $\mathcal{O}(\alpha^5)$ recoil corrections $\Delta_\mathrm{recoil}$ are taken from \cite[Eq.~(14) and (15)]{Antognini:2022xqf} and are consistent with our choice of $F_2$. Since the $F_2$ term in $\Delta_\mathrm{pol}$ cancels exactly with a corresponding term in $\Delta_\mathrm{recoil}$, we do not need to take into account uncertainties of the $F_2$ parametrization in our $\Delta_\mathrm{pol}$ evaluation. Note that radiative corrections (e.g., through electronic vacuum polarization) to the $2\gamma$-exchange diagram are taken into account as well. The proton Zemach radius is then extracted from the $1S$ HFS in H as:
\begin{equation}
    R_\mathrm{Z}=1.036(8)\, \text{fm. }\label{RZprecise}
\end{equation}
This result is more precise than previous extractions \cite{Hyperfine,Antognini:2013rsa}, as can be seen from the top panel of Fig.~\ref{fig:RZ}, where extractions of the Zemach radius from the measured $1S$ HFS in H are shown, assuming the same theoretical prediction of the HFS, but different values of $\Delta_\mathrm{pol}$ as shown in Fig.~\ref{fig:hcompare}. Our evaluation of $\Delta_\mathrm{pol}$ based on new data for the proton spin structure functions (red), can be considered an update of the previous dispersive analysis \cite{Hyperfine} (orange open circle). It has moved closer to the extraction based on LO $\chi$PT (blue), but still does not agree. In the bottom panel of Fig.~\ref{fig:RZ}, recent precise results from lattice QCD suggest a small $R_\mathrm{Z}$ \cite{Djukanovic:2023cqe}, in perfect agreement with the LO $\chi$PT extraction from the 1S H HFS, but in tension with evaluations based on proton form factors measured in scattering \cite{Lin:2021xrc,Distler:2010zq}. Our work is compatible with both  \cite{Djukanovic:2023cqe} and \cite{Distler:2010zq}. 


Given the limited beam time and required tunability of the laser setup, precise theory guidance is crucial for the experiments planned by the CREMA~\cite{Amaro:2021goz} and FAMU~\cite{Pizzolotto:2020fue} collaborations. Here we present an updated theory prediction for the $1S$ HFS in $\mu$H, based on the theory compilation in \cite[Eq.~(40)]{hyperfine_22}, and substituting our $\Delta_\mathrm{pol}$, as well as our extraction of the Zemach radius from the $1S$ HFS in H, cf.~Eq.~(\ref{RZprecise}):
\begin{equation}
E^{\,\text{th.}}_{1S\text{-HFS}}( \mu\mathrm{H})=182.636(16)\,\text{meV}.
\end{equation}
This result is in perfect agreement with the presently most precise prediction presented in \cite[Eq.~(48)]{hyperfine_22} that is 
based on scaling the result from the very accurate electronic hydrogen experimental HFS measurement, cf.\ Eq.~(\ref{expHHFS}), and calculating the small (including small in the uncertainty limits) nonscaling corrections, as suggested in~\cite{Tomalak:2017lxo,Peset:2016wjq}.

We present results for the first ever experimental data in a regime which contributes significantly to the  integrals of the $\Delta_1$ and $\Delta_2$ Hyperfine Splitting contributions. This new data provides previously lacking guidance on how to constrain theoretical calculations of the Hyperfine Splitting effect. Previous data-driven work to determine these quantities~\cite{Hyperfine2} has been limited by older results lacking inelastic proton spin structure function data in the low $Q^2$ regime. These new results are much closer to agreement with $\chi$PT calculations of $\Delta_\mathrm{pol}$, strongly reducing the long-standing tension between methods \cite{hyperfine_22} and reducing the overall error on the polarizability contribution by a factor of two. The reduction in uncertainty provided by our data is crucial in order to facilitate the search for the narrow $\mu$H $1S$ HFS in the planned experiments by the CREMA \cite{Amaro:2021goz} and FAMU \cite{ Pizzolotto:2020fue} Collaborations, as well as to interpret these future measurements. With these experiments aiming at up to 1 ppm relative precision, they have the potential to provide novel insights into the magnetic structure of the proton.

\begin{acknowledgments}
We thank Aldo Antognini and Vadim Lensky for their comments and discussions. This work was supported by
the United States Department of Energy (DOE) 
under award numbers DE-FG02-88ER40410, DE-SC0024665, 
and DE–SC0014434, 
the National Science Foundation under grant PHY-181236,
and by the Schweizerische Nationalfonds.
This work is also supported by the Deutsche Forschungsgemeinschaft (DFG) through the Emmy Noether Programme Grant
449369623, the Research Unit FOR5327 Grant 458854507, and by the Swiss National Science Foundation (SNSF) through the
Ambizione Grant PZ00P2\_193383. This material is based upon work supported by the U.S. Department of Energy, Office of Science, Office of Nuclear Physics under contract DE-AC05-06OR23177.
\end{acknowledgments}

\appendix

\bibliography{g2p_hyperfine.bib}

\end{document}